\documentclass[twocolumn,aps,prl,epsfig,graphics,showpacs]{revtex4}%
\usepackage{graphicx}
\usepackage{amsmath}
\usepackage{amsfonts}
\usepackage{bbm}
\usepackage{amssymb}
\begin{document}
\title{Hybrid Quantum Processors: \\ molecular ensembles as quantum memory for solid state circuits}

\author{P. Rabl$^1$, D. DeMille$^2$, J. M. Doyle$^3$, M. D. Lukin$^3$, R. J. Schoelkopf$^{2,4}$, and P. Zoller$^1$ }
\affiliation{$^1$ Institute for Theoretical Physics, University of
Innsbruck, and Institute for Quantum Optics and Quantum Information
of the Austrian Academy of Sciences, A-6020 Innsbruck, Austria}

\affiliation{$^2$ Department of Physics, Yale University, New Haven, Connecticut 06520, USA}

\affiliation{$^3$ Department of Physics, Harvard University, Cambridge, Massachusetts 02138, USA}

\affiliation{$^4$ Department of Applied Physics, Yale University, New Haven, Connecticut 06520,
USA}

\begin{abstract}
We investigate a hybrid quantum circuit where ensembles of cold polar molecules serve as long-lived
quantum memories and optical interfaces for solid state quantum processors. The quantum memory
realized by collective spin states (ensemble qubit) is coupled to a high-Q stripline cavity via
microwave Raman processes. We show that for convenient trap-surface distances of a few $\mu$m,
strong coupling between the cavity and ensemble qubit can be achieved. We discuss basic quantum
information protocols, including a swap from the cavity photon bus to the molecular quantum memory,
and a deterministic two qubit gate. Finally, we investigate coherence properties  of molecular
ensemble quantum bits.
\end{abstract}


\maketitle

During the last few years we have witnessed remarkable progress  towards realization of  quantum
information processing in various physical systems. Highlights include quantum optical systems of
trapped atoms and ions \cite{Ions2005},  and cavity QED \cite{Cavity}, as well as  solid state
systems including Cooper Pair Boxes (CPB)  \cite{ShnirmanRev,squbits} and quantum dots
\cite{quantumdots2005}. In particular, the strong coupling regime of circuit CQED
\cite{circuitCQED} was realized 
using a CPB strongly coupled to a strip line cavity. In
the light of these developments it is timely to investigate hybrid devices with the goal of
combining the advantages of various implementations, i.e. to build interfaces between, for example,
a quantum optics and solid state qubit with compatible experimental setups
\cite{sorenson04-tian04}. Such interfaces are particularly important in 
 applications where
long-term quantum memories or optical interconnects are required \cite{ftqc,repeater}.

In this Letter we study such a scenario by coupling a strip line cavity to a cloud of cold polar
molecules \cite{Doyle2004_v2}. The cavity may be part of a solid state quantum processor involving
CPB as charge qubits \cite{ShnirmanRev} and microwave photons as a quantum data bus, while the
molecular ensemble serves as a quantum memory with a long coherence time.  The condition of strong
coupling between the cavity and the molecular cloud is achieved via the (large) \emph{electric}
dipole moments of polar molecules for rotational excitations in the electronic and vibrational
groundstate, which are in the tens of GHz regime, and thus provide an ideal match for resonance
frequencies of strip line cavities. By adopting a molecular ensemble instead of a single polar
molecule \cite{single_molec}, we benefit from the enhancement of the coherent coupling $g\sqrt{N}$
with the number of molecules $N$ and $g$ the single molecule vacuum Rabi frequency. This strong
coupling between the molecular ensembles and the circuit CQED system also opens the possibility of
a solid state based readout of molecular qubits. In addition, qubits stored in the molecular
ensemble can be converted to \textquotedblleft flying\textquotedblright\ optical qubits, using
techniques demonstrated for atomic ensembles \cite{AtomicEnsembles}. This provides a natural
interface between mesoscopic quantum circuits and optical quantum communication.

\begin{figure}[ptb]
\begin{center}
\includegraphics[width=0.47\textwidth]{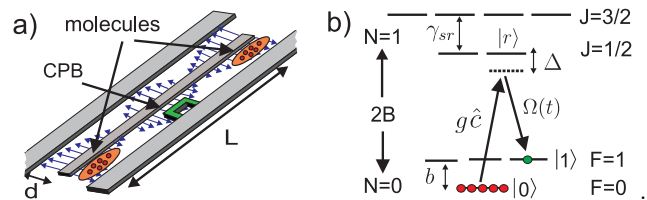} .
\end{center}
\caption{a) 
Ensembles of polar molecules and a CPB are coupled via the quantized field of a strip line cavity
(see text for more details). b) Rotational excitation spectrum of molecules with a
$^{2}\Sigma_{1/2}$ ground state where we plot excited states according to the Hamiltonian
$H=H_{R}+H_{SR}$. For non-zero nuclear spin $\vec I$ (here $I=1/2$) the hyperfine interaction leads
to an additional splitting ($b\sim2\pi \times 100$ MHz) into eigenstates of $\vec F=\vec J+\vec I$
(hyperfine splitting for the excite states is not shown). Two qubit states in the $N=0$ manifold,
$|0\rangle$, $|1\rangle$ are coupled by a Raman process involving a single cavity photon and an
external microwave field,
$\Omega(t)$.}%
\label{fig:Setup}%
\end{figure}

Let us consider the setup of Fig.~\ref{fig:Setup}, where two molecular ensembles are coupled to a
superconducting cavity.
The cavity is assumed to be strongly coupled to a CBP representing a circuit CQED system, as
realized in recent experiments at Yale \cite{circuitCQED}. As discussed in detail below, molecular
spectroscopy allows us to identify longlived states, for example in the form of a spin qubit $|0
\rangle$, $|1\rangle$ in the ground rotational manifold.
Starting with a cloud of $N$
molecules prepared in $|0\rangle_m\equiv|0_{1}0_{2}%
\dots0_{N}\rangle$ coupling to a microwave or cavity field leads to excitations in the form of
symmetric Dicke states, $|1\rangle_m\equiv1/\sqrt {N}\sum_{i}|0_{1}..1_{i}..0_{N}\rangle\equiv
m^{\dag}|0\rangle_m$ \emph{etc.} For weak excitation the operator $m$ obeys approximate harmonic
oscillator commutation relations $[m,m^{\dag}]\approx1$, and the ensemble excitations are
conveniently described as a set of harmonic oscillator states $|0\rangle_m$, $|1\rangle_m\equiv
m^{\dag}|0\rangle_m$ \emph{etc.} Our goal below is to use the lowest two of these states as
ensemble qubits, which can be manipulated by coupling them to the superconducting cavity and a CPB.

The dynamics of the coupled system (Fig.~1a) can be described in terms of a Hamiltonian
$H_{\mathrm{sys}}=H_{C}+H_{M}+H_{CM}$, which is the sum of a Jaynes-Cummings type Hamiltonian for
the circuit CQED system $H_{C}$, a Hamiltonian for the (spin) excitations of the molecular
ensembles $H_{M}$, and the coupling of the molecules to the cavity $H_{CM}$. The CQED Hamiltonian
has the form
\begin{equation}
\label{eq:Hsys}H_{C}=-\delta_{c}(t)|e\rangle\langle e|+g_{c}(|e\rangle\langle
g|\hat{c}+|g\rangle\langle e|\hat{c}^{\dag})\,.
\end{equation}
Here $|g\rangle$ and $|e\rangle$ denote the ground and the first excited eigenstate of the CPB at
the charge degeneracy point representing a charge qubit with a (tunable) transition frequency\
$\omega_{cq}(t)$ with detuning $\delta_{c}(t)=\omega_{c}-\omega_{cq}(t)$ from the cavity frequency
$\omega_{c}$. The operator $\hat{c}$ ($\hat{c}^{\dag}$) is the cavity annihilation (creation)
operator for microwave photons of frequency $\omega_{c}$, and $g_{c}$ is the vacuum Rabi
frequency.\ Eq.~\eqref{eq:Hsys} is written in an interaction picture with the bare cavity
Hamiltonian $\omega _{c}\hat{c}^{\dag}\hat{c}$ transformed away. The Hamiltonian describing the
internal excitations of the molecular ensembles $i=1,2$ and the coupling of
ensemble states to the stripline cavity takes the form%
\begin{equation}
H_{M}+H_{CM}=-\sum_{i}\delta_{m}^{(i)}(t)\,m_{i}^{\dag}m_{i}+\sum_{i}%
g_{m}^{(i)}(t)\,m_{i}^{\dag}\hat{c}+\mathrm{H.c.}\nonumber
\end{equation}
Before we enter the details of the derivation of $H_{M}$ and $H_{CM}$
we note the basic structure of the Hamiltonian $H_{\mathrm{sys}}$. The ensemble excitations and the
cavity represent a system of coupled harmonic oscillators interacting with a two-level system (CPB)
with controllable coefficients. In general, this provides the basic ingredients for (i) swap
operations between the charge, cavity and ensemble qubits, (ii) single qubit rotations of the
molecular qubit via the charge qubit, and (iii) 2-qubit entanglement operations between two
ensemble qubits via the cavity mode, where the charge qubit plays the role of a nonlinearity. For
example, the CPB can act as a \textquotedblleft single photon source\textquotedblright, i.e. we
generate a superposition state of the charge qubit, which by an appropriate control sequence can be
swapped over to the cavity, and is finally stored in one of the molecular ensembles, and vice
versa: $\left( \alpha|g\rangle+\beta|e\rangle\right)
|0\rangle_c|0\rangle_{m}\rightarrow|g\rangle\left(\alpha |0\rangle_{c}+\beta|1\rangle_{c}\right)
|0\rangle_{m}\rightarrow |g\rangle|0\rangle_{c}\left( \alpha|0\rangle_{m}+\beta|1\rangle
_{m}\right)  $.

The above discussion has ignored various sources of decoherence. In the Yale experiment
\cite{circuitCQED}, the circuit CQED system realizes the strong coupling regime with vacuum Rabi
frequency $g_{c}\lesssim 2\pi\times 50$ MHz. The decoherence of the charge qubit is dominated by
the dephasing rate $T_2^{-1}\approx2\pi\times 0.5$ MHz, while the photon loss rate is
$\kappa/2\pi=1$ to $0.01$ MHz, i.e. the charge qubit is the dominant source of decoherence. Below
we will show that for a cloud of $N\approx10^{4}$ to $10^{6}$ molecules of temperature $1$ mK,
trapped at a distance $10\,\mu$m above the strip line cavity, one can reach the regime of strong
cavity - ensemble coupling $g_{m}/2\pi\approx$ $1$ to $10$ MHz, which should be compared with the
expected collisional dephasing rates of the molecular memory of a few hundred Hz.

Fig. 1b shows the rotational spectrum of CaF, which provides an example for spectra of
alkaline-earth monohalogenides with a $^{2}\Sigma_{1/2}$ ground state corresponding to a single
electron outside a closed shell.
The spectrum consists of rotational eigenstates, described by a rigid rotor Hamiltonian
$H_{R}=B\vec{N}^{2}$ with $B\sim2\pi\times 10$ GHz the rotational constant, and $\vec{N}$ the
angular momentum of the nuclei. The unpaired spin is coupled to the molecule rotation according to
$H_{SR}=\gamma_{sr}\vec{S}\vec{N}$ with $\gamma_{sr}\sim 2\pi\times 40$ MHz and $\vec{S}$
the electron spin ($S=1/2$). Coupled eigenstates are denoted by $|N,S,J;M_{J}%
\rangle$ with $\vec{J}=\vec{N}+\vec{S}$. As seen from Fig.~\ref{fig:Setup}, there is a spin
rotation splitting ($\rho$-doubling) for rotationally excited states. In addition, there can be
hyperfine interactions, which in particular lead to a splitting of the ground state $N=0$, as in
the case of CaF with a nuclear spin $I=1/2$ which are coupled with $J=1/2$ to $F=0$ and $1$ states.
In the following, we denote by $|0\rangle$, $|1\rangle$ a pair of states in the rotational ground
state manifold to provide our spin qubit. 
Compared to qubits
stored in the rotational degrees of freedom this choice of states avoids unfavourable
$N=1\rightarrow N=0$ collisions while $|0\rangle$ and $|1\rangle$ can still be coupled efficiently
by a Raman process.


The cavity mode and microwave fields of appropriate frequency and polarization couple rotational
ground states to excited states with electric dipole matrix elements $\mu$ ($\sim$ $5$ Debye). Two
microwave
driving fields provide an effective coupling Hamiltonian $\frac{1}{2}%
\Omega_{\text{\textrm{eff}}}(t)|0\rangle\langle 1|\!+\mathrm{h.c.}$ to rotate the single molecule
spin qubit, where 
$\Omega_{\text{\textrm{eff}}}=\Omega_{1}\Omega_{2}/2\Delta$, with $\Omega_{1,2}$ the Rabi
frequencies and $\Delta$ the detuning from the excited state $|r\rangle$,
($\Delta\gtrsim\Omega_{1,2}$). By similar arguments the coupling to the cavity has the form
$g_{\text{\textrm{eff}}}(t)|1
\rangle\langle0|\hat{c}+\mathrm{h.c.}$ with $g_{\text{\textrm{eff}%
}}(t)\equiv g\Omega(t)/2\Delta$, where $g=\mu\mathcal{E}_{c}$ is the
vacuum Rabi frequency, and $\mathcal{E}_{c}\approx\sqrt{\hbar\omega_{c}%
/2\pi\epsilon_{0}d^{2}L}$ is the electric field per photon for a cavity length $L$ and typical
electrode distance $d$. Typical values are $g/2\pi\sim$ $5-10$ kHz for $\mu\approx5$ D and
$d\approx10\,\mu\mathrm{m}$. The distance $d$ is also an estimate of the trapping distance of the
molecular cloud from the cavity, which is well in the limit where standard trapping techniques work
reliably and surface effects are negligible. Rewriting $H_{CM}$ in terms of the collective operator
$m$ we obtain an effective cavity - ensemble coupling $g_{m}(t)=\sqrt{N}g_{\mathrm{eff}}(t)$. Due
to the large wavelength $\lambda_{c}\approx1.5$ cm a trap volume of $V\leq d\times d\times\lambda
_{c}/10$ contains $N=10^{4}\dots10^{6}$ molecules for gas densities of
$n\sim10^{12}\,\mathrm{cm}^{-3}$ resulting in a coupling strength of $g_{m}/2\pi \sim1$ -- $10$
MHz. The parameters $\delta^{(i)}_{m}(t)$ are Raman detunings, which can be controlled
independently, e.g. by applying local magnetic and/or electric fields.  Thus we obtain the
Hamiltonian\ $H_{M}+H_{CM}$, which allows a SWAP of a cavity and an ensemble state, for example, by
an adiabatic sweep of $\delta_{m}$ across the resonance. This
corresponds to a read / write operation $\rho_{c}\otimes|0\rangle_{m}%
\langle0|\longleftrightarrow|0\rangle_{c}\langle0|\otimes\rho_{m}$ with $\rho_{c}$ an arbitrary
density operator of the microwave field in the cavity, and $\rho_{m}$ the identical state stored in
ensemble excitations.

\begin{figure}[ptb]
\begin{center}
\includegraphics[width=0.47\textwidth]{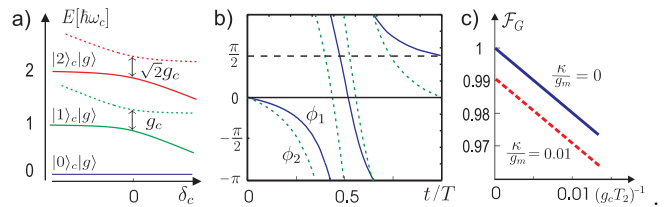} .
\end{center}
\caption{a) Adiabatic energies levels for the states $|n\rangle_{c}|g\rangle$
as a function of the charge qubit detuning $\delta_{c}$. b) Evolution of the
dynamical phases $\phi_{1}$ (solid line) and $\phi_{2}$ (dashed line) for a
pulse $\delta_{c}(t)=-\delta_{0}(2t/T-1)^{2}-\delta_{1}$ , $\delta_{0}%
/g_{c}=30$, $\delta_{1}/g_{c}=0.44$ and $T=44.79/g_{c}$. For the same pulse
$\delta_c(t)$ the resulting fidelity of the total gate sequence, $\mathcal{F}%
_{G}$ (averaged over all initial states $|\psi\rangle_{m}$), is plotted in c) for different values
of the charge qubit dephasing rate, $T_2^{-1}$, and the cavity loss rate $\kappa$.
}%
\label{fig:TwoQubitGate}%
\end{figure}

The CPB provides a nonlinear element in the Hamiltonian $H_{\mathrm{sys}}$.
This allows first of all single qubit operations of ensemble qubits, e.g. by combining a swap
operation with single qubit rotations of the charge qubit, and second, deterministic entanglement
operations of qubits stored in two molecular ensembles. An example of such a protocol, which uses
the CPB as a nonlinear phase shifter, is given as follows. We assume that the system is initially
prepared in the state $|\psi\rangle_{t=0}=|\psi\rangle_{m} |0\rangle_{c}|g\rangle$ with the charge
qubit far detuned from the cavity resonance, $|\delta_{c}(0)|\gg g_{c}$, and the two ensemble
qubits in an
arbitrary state $|\psi\rangle_{m}$ spanned by the basis $|\epsilon_{1}%
\epsilon_{2}\rangle_{m}$, $\epsilon_{i}=0,1$. In a first step, in analogy to the single qubit swap,
the state $|\psi\rangle_{m}$ is (partially) transferred to the cavity. Assuming symmetric
conditions, $g_{m}^{(1)}=g_{m}^{(2)}$ and $\delta_{m}^{(1)}(t)=\delta_{m}^{(2)}(t)$ it is
convenient to rewrite the ensemble state in terms of the (anti)symmetric operators
$m_{s/a}=(m_{1}\pm m_{2})/\sqrt{2}$ acting on $|00\rangle_{m}$. An adiabatic sweep of the Raman
detunings then realizes the swap operation
$m_{s}^{\dag}|00\rangle_{m}|0\rangle_{c}\rightarrow|00\rangle_{m}
|1\rangle_{c}$ and $(m_{s}^{\dag})^{2}|00\rangle_{m}|0\rangle_{c}%
\rightarrow\sqrt{2}|00\rangle_{m}|2\rangle_{c}$ while the states $|00\rangle_{m}|0\rangle_{c}$,
$m_{a}^{\dag}|00\rangle_{m}|0\rangle_{c}$ and $(m_{a}^{\dag})^{2}|00\rangle_{m} |0\rangle_{c}$
remain unaffected. In a second step, the charge qubit is adiabatically tuned close to resonance for
a time $T$, $|\delta_{c}(T/2)|\lesssim g_{c}$. During this pulse the non-vacuum states acquire a
nonlinear dynamical phase, $|n\rangle_{c}\rightarrow e^{i\phi_{n}}|n\rangle_{c}$ with
$\phi_{n}=-\int_{0}^{T}dt^{\prime}(\delta _{c}(t^{\prime})+\sqrt{\delta_{c}^{2}(t^{\prime})+n4
g_{c}^{2}}\,)/2$. The pulse form $\delta_{c}(t)$ and the length $T$ are chosen such that $\phi
_{1}\simeq\pi/2$ and $\phi_{2}=2\pi n$ (see e.g. Fig.~\ref{fig:TwoQubitGate}). The second condition
ensures that after writing the cavity state back into the ensembles states, i.e. reversing step
one, the ensemble states $|20\rangle _{m}$ and $|02\rangle_{m}$ remain unpopulated. The total gate
sequence corresponds to a \textquotedblleft$\sqrt{\mathrm{SWAP}}$"-like gate for two ensembles
qubits, with $|00\rangle_{m} \rightarrow|00\rangle_{m}$,
$|10\rangle_{m} \rightarrow e^{i\pi/4}(|10\rangle_{m}+i|01\rangle_{m}%
)/\sqrt{2}$, $|01\rangle_{m} \rightarrow e^{i\pi/4}(i|10\rangle_{m}%
+|01\rangle_{m})/\sqrt{2}$ and $|11\rangle_{m} \rightarrow|11\rangle_{m}$. A numerical simulation
of this gate sequence based on a master equation treatment of the dissipative terms
\cite{circuitCQED} shows that the gate fidelity is only limited by $(g_cT_2)^{-1}$ i.e. the
decoherence of the CPB during the time it is tuned close to resonance (see Fig.
\ref{fig:TwoQubitGate} for more details).

We now turn to an analysis of decoherence in the molecular ensemble. In particular collisional
dephasing of the ensemble qubit, and a spatial variation of the cavity-molecule coupling $g_{\rm
eff}(x)$ in combination with the thermal motion of the molecules in the trap contribute to a finite
decoherence time of the molecular quantum memory, and result in imperfections during gate
operations.

An operational definition of the decoherence time of the ensemble qubit can be given in terms of an
(idealized) experiment. A cavity qubit $\rho _{c}(t=0)=|\psi\rangle_{c}\langle\psi|$ with
$|\psi\rangle_{c}=\alpha |0\rangle_{c}+\beta|1\rangle_{c}$ is written at time $t=0$ to the
molecular memory with all molecules initialized in the state $|0\rangle$ in a (perfect) swap
operation, kept in storage for a time interval $\tau$, undergoing dephasing collisions. At
$t=\tau$, the qubit is transferred back to the cavity mode, resulting in a reduced density matrix
$\rho_{c}(\tau)$ of the cavity with a fidelity
$F_{\tau}=\min_{\psi_{c}}\,_{c}\langle\psi|\rho_{c}(\tau)|\psi\rangle_{c}$, with the decoherence
time of the ensemble memory identified as the decay time of the fidelity. The analysis of this
process resembles the discussions of clock shifts, and in particular studies of collisional
dephasing of spins in thermal and quantum degenerate atomic clouds in a Ramsey interferometry
setup.  A formal theoretical description of these phenomena is provided by quantum kinetic theory
\cite{QKT}.

We consider a thermal cloud of molecules in the lowest rotational state with qubits stored in spin
or hyperfine states (Fig.~\ref{fig:Setup}). Molecules in the rotational ground state are trapped
magnetically \cite{mtrap} or by a (spin independent) electric rf-trap \cite{rftraps}. In a magnetic
trap two molecules interact asymptotically according to a $V(r)\simeq-C_{6}/r^{6}$ potential with
$C_{6}= (\mu^{2}/4\pi\epsilon_0)^2/6B$. The effective range of this potential is given by
$R_{\ast}=\sqrt[4]{mC_{6}/\hbar^{2}}$, which provides an estimate for the s-wave scattering length
$\bar{a}$. For example, for CaCl (which has two magnetically trapped hyperfine states) we obtain
$R_{\ast}\approx$ 780 $a_{B}$ which is a few times the typical scattering length encountered for
alkali atoms. S-wave scattering dominates for temperatures $T\lesssim T_*\approx 1\mu$\textrm{K}
where the thermal energy is below the centrifugal barrier for higher angular momenta, leading to an
estimate for the collision rate $\gamma_{\mathrm{col}}=8\pi \bar{a}^{2}n\bar{v}\approx 2\pi\times
150$ Hz for $n=10^{12}\mathrm{cm}^{-3}$ and $\bar{v}$ the relative thermal velocity. For
temperatures $T\gg T_{\ast}$ higher partial waves contribute, and an estimate of the cross section
based on the unitarity limit gives $\gamma_{\mathrm{col}}\lesssim 2\pi\times$700 Hz for $T=1$ mK.
In electric traps the induced dipole moments, $\mu_{\rm ind}$, lead to a $V(r)\sim \mu_{\rm
ind}^2/r^3$ dependence of the asymptotic interaction. Although this long-range behavior
significantly changes the low temperature scattering ($T < 1\mu{\rm K}$) the estimate based on the
unitarity limit at $T\approx 1$ mK still provides a valid bound for scattering rates of weakly
polarized molecules ($\mu_{\rm ind} < 1$D).

We have calculated the decoherence time of an ensemble qubit corresponding to collisional dephasing
using quantum kinetic theory \cite{unpublished}. Dephasing of the qubit coherence
$(\rho_{c})_{10}(\tau)=\exp(-\gamma_{10}\tau/2)(\rho _{c})_{10}(0)$ is associated with spin
dependent collisions between the states $|0\rangle$ and $|1\rangle$, which gives a contribution
\begin{align*}
\gamma_{10}  & =\frac{2\hbar^{3}n}{m^{2}}\int\prod_{i=1}^{4}d^{3}k_{i}%
\,\delta(\Delta E(\mathbf{K}))\,\delta(\mathbf{K})\,P(\vec{k}_{1})P(\vec
{k}_{2})\\
& \times (|f_{00}^{e}(\mathbf{K})-f_{01}^{e}(\mathbf{K})|^{2}+ |f_{00}^{in}({\bf K})|^2 +
|f_{01}^{in}({\bf K})|^2)\,.
\end{align*}
It depends on the difference between $f_{00}^{e}$ and $f_{01}^{e}$, the elastic scattering
amplitudes for the internal states $|00\rangle$ and $(|10\rangle+|01\rangle)/\sqrt {2}$ averaged
over the thermal distributions $P(\vec{k})$ in the scattering process between momenta
$\mathbf{K}=(\vec{k}_{3},\vec{k}_{4}\!\leftarrow \!\vec{k}_{1},\vec{k}_{2})$, and
$\delta$-functions accounting for energy and momentum conservation in the collision. In addition,
there may be contributions from inelastic collisions, $f_{00}^{in}$ and $f_{00}^{in}$, which
scatter molecules outside the $|0\rangle$, $|1\rangle$ subspace. While accurate scattering
amplitudes for molecular collisions may not be available at present, we can estimate these
contributions in certain limits.
For $s$-wave scattering the above expression simplifies to $\gamma_{10}%
=8\pi(a_{00}-a_{01})^{2}n\bar{v}$ with $a_{00}$ and $a_{01}$ scattering lengths. If we assume that
the scattering length is dominated by a
spin exchange potential, the scattering is characterized by a singlet ($a_{S}%
$) and triplet scattering length ($a_{T}$). In the simple case of a pure spin qubit
$\{|0\rangle,|1\rangle\}\equiv\{|S\!=\!1/2,m_{s}\!=\!\pm 1/2\rangle\}$ we find
$a_{00}=a_{01}=a_{T}$, and the dephasing rate is determined by non-vanishing contributions arising
from magnetic dipole and spin rotation coupling, which are expected to be much smaller. In a
similar way, in the presence of hyperfine interactions we can form a qubit
$|0\rangle=|F\!=\!I\!+\!1/2,M_{F}\!=\!F\rangle$ and $|1\rangle
=|F'\!=\!I\!-\!1/2,M_{F}\!=\!F'\rangle$, where again  $|00\rangle$ and $|01\rangle+|10\rangle$
contain no spin singlet contribution and the leading dephasing term vanishes. In the worst case the
decoherence rate is bounded by the single molecule collision rate
$\gamma_{10}\approx\gamma_{\mathrm{col}}$ which has been estimated above.


Spatial variations of the effective single molecule-cavity coupling, $ g_{\rm eff}(x)$, result in a
dephasing of the qubit during a single swap gate and an incomplete recovery of the state after a
redistribution of the molecules between two successive write/read operations. The inhomogeneity in
the coupling arises from the variation of the cavity mode function on a scale of the electrode
distance, $g(x)\approx g(1-\alpha x/d)$, with $\alpha$ a numerical constant, and a position
dependence of the detuning $\Delta(x)\approx \Delta- m\delta \omega^2 x^2/(2\hbar)$. Here $\delta
\omega^2=\omega_t^2-\omega_{r}^2$ accounts for a
 difference in the trapping potentials for the qubit states ($\omega_t$) and the excited
state $|r\rangle$ ($\omega_{r}$). For an optimal detuning $\Delta_*\simeq\sqrt[3]{3
g^2N(k_bT\delta\omega^2)^2/\kappa\omega_t^4\hbar^2}$ the inhomogeneous coupling results in a total
gate error of $\epsilon\approx \alpha^2(k_bT/m\omega_t^2d^2)+(k_bT\delta\omega^2 \kappa/\hbar
g^2N\omega_t^2)^{2/3}$ \cite{unpublished}. For $g\sqrt{N} \sim 2\pi\times 10$ MHz, $\kappa \sim
2\pi\times 10$ kHz and
 at $T=1$ mK gate fidelities of $\mathcal{F}>0.99$ require trap frequencies of $\omega_{t}\sim
2\pi\times 50$ kHz and a similar trapping potential for the state $|r\rangle$ ($\delta \omega^2
\sim 0.1 \omega^2_t$). Lower temperatures and an optimized cavity / trap design, e.g.
$\alpha,\delta \omega^2 \rightarrow 0$, lead to a further significant reduction of gate errors.

In conclusion, thermal ensembles of cold polar molecules represent a
good quantum memory that can be strongly coupled to strip line cavities with long
life time, limited essentially only by collisional dephasing. We
note that these dephasing channels can be virtually eliminated, if
the ensemble is prepared in a high-density crystalline phase of
dipolar gases with dipole moments induced and aligned by a DC
electric field under 2D trapping conditions \cite{unpublished}. The
present work opens an exciting avenue towards long
lived molecular quantum memory for solid state quantum processors.

\begin{acknowledgments}
Work at Innsbruck is supported by the Austrian Science Foundation, European Networks and the
Institute for Quantum Information. P.R. thanks the Harvard Physics Department and ITAMP for
hospitality. Work at Harvard is supported by NSF, Harvard-MIT CUA and Packard and Sloan
Foundations. Work at Yale is supported by NSF Grant DMR0325580, the W.M. Keck Foundation, and the
Army Research Office.
\end{acknowledgments}


\begin{thebibliography}{99}
\bibitem{mtrap} Note that magnetic trapping achieved using localized magnetic fields near
the nodes of the strip-line resonator should not significantly degrade the quality of the
superconducting cavity.
\end{thebibliography}

\end{document}